\documentclass[nofootinbib,twocolumn,prd,superscriptaddress,preprintnumbers,prd]{revtex4-1}

\usepackage{slashed,bm,graphicx}
\usepackage{amssymb}
\usepackage{amsmath}
\usepackage{mathrsfs}
\usepackage{mathtools}
\usepackage{color}

\def\nn{\nonumber \\}

\def\vev#1{\left\langle #1 \right\rangle }
\def\abs#1{\left| #1 \right| }
\def\tr{ \text{Tr}\, }

\def\rd{ {\rm d}}
\def\ip#1#2{ \left\langle #1, #2 \right\rangle}
\def\msm{S^3}
\def\m{ \mathcal{M} }
\def\g{ \mathcal{G} }
\def\h{ \mathcal{H} }

\begin{document}

\preprint{CERN-TH-2016-024}

\title{Sigma Models with Negative Curvature}

\author{Rodrigo Alonso}

\affiliation{\vspace{1mm}
Department of Physics, University of California at San Diego, La Jolla, CA 92093, USA}

\author{Elizabeth E.~Jenkins}
\author{Aneesh V.~Manohar}

\affiliation{\vspace{1mm}
Department of Physics, University of California at San Diego, La Jolla, CA 92093, USA}

\affiliation{\vspace{1mm} CERN TH Division, CH-1211 Geneva 23, Switzerland}

\begin{abstract}
We construct Higgs Effective Field Theory (HEFT) based on the scalar manifold $\mathbb{H}^n$, which is a hyperbolic space of constant negative curvature. The Lagrangian has a non-compact $O(n,1)$ global symmetry group, but it gives a unitary  theory as long as only a compact subgroup of the global symmetry is gauged. Whether the HEFT manifold has positive or negative curvature can be tested by measuring the $S$-parameter, and the cross sections for longitudinal gauge boson and Higgs boson scattering, since the curvature (including its sign) determines deviations from Standard Model values.
\end{abstract}
\maketitle

%------------------------------------------------------------------------------
\section{Introduction}
%------------------------------------------------------------------------------

The recently discovered neutral scalar particle with a mass of $\sim 125$\,GeV has led to renewed interest in models of electroweak symmetry breaking. The Standard Model (SM) is one such theory, where the electroweak symmetry is broken by a complex scalar doublet $H$ that transforms linearly as $\mathbf{2}_{1/2}$ under the $SU(2)_L \times U(1)_Y$ gauge symmetry. Generalizations of the SM include the Standard Model Effective Field Theory (SMEFT), which is the SM plus higher dimension operators, and Higgs Effective Field Theory (HEFT)~\cite{Feruglio:1992wf,Grinstein:2007iv}, which contains the three ``eaten'' Goldstone boson degrees of freedom of the SM in a chiral field $\xi(x)$, and an additional neutral scalar degree of freedom $h(x)$. The geometry of the scalar manifold $\m$ is an interesting object that can be studied experimentally, as discussed in  Ref.~\cite{Alonso:2015fsp}. In the SM, the scalar manifold $\m$ is flat. Deviations from the SM cross section for Higgs boson and longitudinal $W^\pm, Z$ gauge boson scattering are proportional to the curvature. In particular, the sign of the deviation depends on whether the 
curvature of $\m$ is positive or negative. In composite Higgs models~\cite{Kaplan:1983fs}, the Higgs field is a pseudo-Goldstone boson generated by the symmetry breaking $\g \to \h$ of a compact group $\g$, and the scalar manifold is $\m= \g / \h $ which has positive curvature. HEFT is more general, and can accomodate manifolds with any curvature. In this paper, we give a simple example of a sigma model where $\m$ has negative curvature.

\begin{figure}
\centering
\includegraphics[]{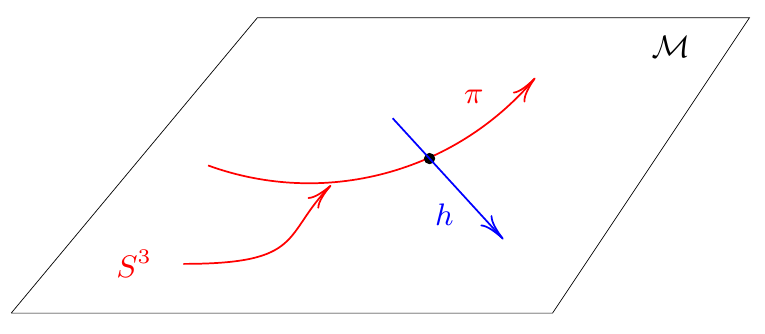}
\caption{\label{fig:1} The scalar manifold $\m$ for HEFT. The space breaks up into angular coordinates $\pi^a$ which form
the (eaten) Goldstone boson manifold $\msm$ of the SM, and a radial direction $h$. }
\end{figure}

Custodial $SU(2)$ symmetry plays an important role in the SM, and we will assume that it is also a symmetry in the sigma model, so that the electroweak symmetry breaking pattern is $O(4) \to O(3)$. The group $O(4) \sim SU(2)_L \times SU(2)_R$, where $SU(2)_L$ is weak $SU(2)$, and the $T_3$ generator of $SU(2)_R$ is weak hypercharge.
Schematically, the scalar manifold $\m$ of HEFT is shown in Fig.~\ref{fig:1}. The angular directions live on $O(4)/O(3) \sim S^3$, and are the three Goldstone bosons eaten by the Higgs mechanism. There is one (or more) additional scalar field direction $h$, often referred to as the radial direction.

%%
%% FIG 1
%%
%%
%% END FIG 1
%%
The SM and SMEFT are special cases of HEFT, as can be seen by using the exponential parameterization
\begin{align}
H &
=\frac{1}{\sqrt 2} \left[ \begin{array}{cc}  i\varphi_1+\varphi_2 \\ \varphi_4 - i \varphi_3  \end{array}\right] = \frac{1}{\sqrt 2}(v+h) e^{i \bm{\tau \cdot \pi}/v}\left[ \begin{array}{cc} 0 \\ 1  \end{array}\right] \nn
&\equiv \frac{1}{\sqrt 2}(v+h) \,{\xi} \left[ \begin{array}{cc} 0 \\ 1  \end{array}\right]\,.
\label{1}
\end{align}
The surface spanned by the angular coordinates $\bm{\pi}$ is the three-sphere $S^3$ of radius $v$,
\begin{align}
\varphi_1^2 + \varphi_2^2+\varphi_3^2+\varphi_4^2 &=v^2\,,
\end{align}
which is the vacuum manifold of the SM. The radius of the sphere, $v\sim 246$ GeV, is fixed by the gauge boson masses.

The manifold structure of HEFT,  assuming custodial symmetry, contains the coset space
\begin{align}
\m \supset \msm = O(4)/O(3)\,,
\end{align}
where $\m$ has field coordinates $h,{\xi}$, with ${\xi}$ a function of the  Goldstone bosons $\bm{\pi}$. The angular scalar fields $\bm{\pi}$ are eaten by the Higgs mechanism, producing the longitudinal polarization states of the massive $W^\pm$ and $Z$ bosons. $\msm$ is a submanifold of $\m$ given by fixing $h$. Each value of $h$ gives an $\msm$, which  together form a foliation of $\m$, in the same way that a sequence of concentric spheres gives a foliation of $\mathbb{R}^n$.

Using polar coordinates for the SM,
\begin{align}
\bm{\varphi} &= (v+h) \bm{n}\,,
\end{align}
where $\bm{n}$ is a four-dimensional unit vector, $\bm{n} \in S^3$, gives the scalar kinetic term
\begin{align}
L &= \frac12 \left(\partial_\mu h \right)^2 + \frac12 (v+h)^2 \left(\partial_\mu \bm{n} \right)^2 \,.
\label{5a}
\end{align}
The HEFT kinetic term is the generalization of Eq.~(\ref{5a}),
\begin{align}
L &= \frac12 \left(\partial_\mu h \right)^2 + \frac12 F(h)^2 v^2 \left(\partial_\mu \bm{n} \right)^2 
\label{heft}
\end{align}
where $F(h)$ is an arbitrary radial function.  The HEFT radial function satisfies
\begin{align}
F(0)=1
\label{f0}
\end{align}
since the radius of $S^3$ is fixed to be $v$.
In the SM, the radial function is
\begin{align}
F(h) &= \left(1 + \frac{h}{v} \right)^2\,.
\label{fsm}
\end{align}

%------------------------------------------------------------------------------
\section{The $O(5) \to O(4)$ model}\label{sec:s4}
%------------------------------------------------------------------------------

We start by discussing the well-known $O(5) \to O(4)$ composite Higgs model~\cite{Agashe:2004rs}. This example is the simplest composite Higgs model incorporating custodial $SU(2)$ symmetry. The presentation of this model introduces the notation and formalism we will use, which carries over to the negative curvature case with only a few crucial sign changes.

\begin{figure}
\centering
\includegraphics[]{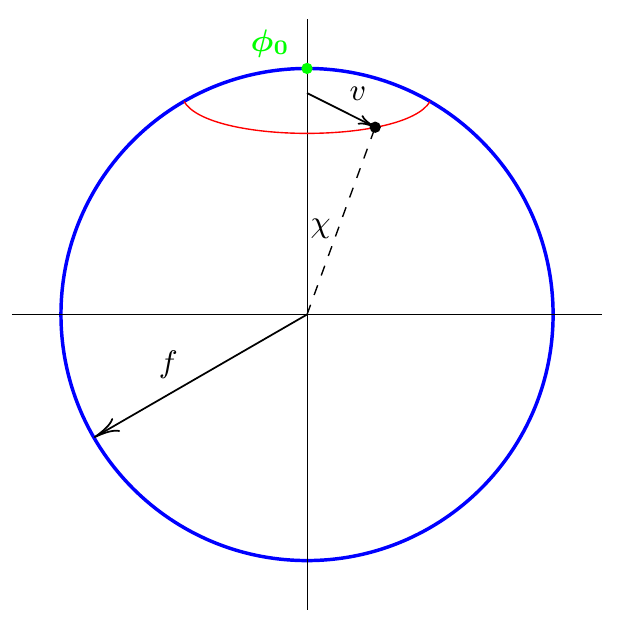}
\caption{\label{fig:2} The scalar manifold of the $O(5) \to O(4)$ sigma model is the sphere $S^4$ of radius $f$. The SM $SU(2) \times U(1)$ gauge group is a subgroup of the unbroken group $\h$ that leaves $\bm{\phi}_0$ invariant. The four-component field $\bm{\varphi}$ gets a VEV $v\ll f$, which breaks electroweak symmetry. }
\end{figure}

Consider a five-dimensional scalar field $\bm{\phi}$ which lives on a flat scalar manifold $\m \sim \mathbb{R}^5$ which has an $O(5)$ symmetry.  The $O(5)$ generators are
\begin{align}
i \left[ T^{ab}  \right]_{ij} &=  \delta_{ai} \delta_{bj} - \delta_{aj} \delta_{bi}\,
\end{align}
with $1 \le a < b \le 5$, so that
\begin{align}
i \, T^{12}   &=  \left[ \begin{array}{ccccc} 
0 & 1 & 0 & 0 & 0 \\ 
-1 & 0 & 0 & 0 & 0 \\ 
0 & 0 & 0 & 0 & 0 \\ 
0 & 0 & 0 & 0 & 0 \\ 
0 & 0 & 0 & 0 & 0 \\ 
\end{array}\right],\ \hbox{etc.}
\label{5}
\end{align}
The theory has a potential $V(\bm{\phi \cdot \phi})$ with a minimum at $\bm{\phi \cdot \phi} = f^2$. 
The vacuum manifold of the theory is the sphere $S^4$ of radius $f$, as shown in Fig.~\ref{fig:2}. The choice of vacuum expectation value
\begin{align}
\vev{ \bm{\phi}_0} &= \left[ \begin{array}{c} 0 \\ 0 \\ 0 \\ 0 \\ f \end{array}\right]
\end{align}
breaks the $O(5)$ symmetry to $O(4)$, giving four Goldstone bosons.
%%%
%%% BEGIN FIG 2
%%%
%%%
%%% END FIG 2
%%%

The four broken generators are $T^{a5}$, $a=1,2,3,4$, where
\begin{align}
i \omega^a T^{a5} &= \left[ \begin{array}{cccc|c} 
0 & 0 & 0 & 0 & \omega^1 \\ 
0 & 0 & 0 & 0 & \omega^2 \\ 
0 & 0 & 0 & 0 & \omega^3 \\ 
0 & 0 & 0 & 0 & \omega^4 \\ 
\hline
-\omega^1 &  -\omega^2 &  -\omega^3 &  -\omega^4 & 0 \\ 
\end{array}\right]\,.
\label{7}
\end{align}
The unbroken $O(4)$ Lie algebra is isomorphic to $SU(2)_L \times SU(2)_R$, which is generated by
\begin{align}
\mathbf{T}_L &= \frac12 (\mathbf{J}+\mathbf{K})\,, &
\mathbf{T}_R &= \frac12 (\mathbf{J}-\mathbf{K})\,,
\end{align}
where $\mathbf{J}^a = \epsilon_{abc} T^{bc}$, $\mathbf{K}^a = T^{a4}$, and
\renewcommand{\arraystretch}{1.2}
\begin{align}
i A_L \cdot T_L &= \frac12 \left[ \begin{array}{cccc|c} 
0 & A_L^3 & -A_L^2 & A_L^1 & 0 \\
-A_L^3 & 0  & A_L^1 & A_L^2  & 0 \\
A_L^2 & -A_L^1 & 0 & A_L^3 & 0 \\
-A_L^1 & -A_L^2 & -A_L^3 & 0 & 0 \\
\hline
0 & 0 & 0 & 0 & 0
\end{array} \right] \,, \nn[10pt]
i A_R \cdot T_R &= \frac12 \left[ \begin{array}{cccc|c} 
0 & A_R^3 & -A_R^2 & -A_R^1 & 0\\
-A_R^3 & 0  & A_R^1 & - A_R^2 & 0\\
A_R^2 & -A_R^1 & 0 & - A_R^3 & 0 \\
A_R^1 & A_R^2 & A_R^3 & 0 & 0\\
\hline
0 & 0 & 0 & 0 & 0
\end{array} \right]\,,
\label{8p}
\end{align}
which will be used to construct the gauge covariant derivative.

A general point of $\m$ in the Northern hemisphere can be parameterized as
\begin{align}
\bm{\phi} &= \left[ \begin{array}{c} f \sin \chi\, \bm{n} \\ f \cos \chi \end{array}\right]\,,
\label{15}
\end{align}
where $\bm{n}$ is a four-dimensional unit vector, which transforms linearly as the four-dimensional representation under the unbroken $O(4)$ symmetry. $\chi$ and $\bm{n}$ together have four degrees of freedom.
An alternate square-root parameterization is also useful,
\begin{align}
\bm{\phi} &= \left[ \begin{array}{c} \bm{\varphi} \\[5pt] \sqrt{f^2- \bm{\varphi \cdot \varphi}} \end{array}\right]\,,
\end{align}
where $\bm{\varphi}$ has 4 components.
The kinetic term of the composite Higgs theory is
\begin{align}
L &=
\frac12\partial_\mu \bm{\phi} \cdot \partial^\mu \bm{\phi} = \frac12 f^2 \left(\partial_\mu \chi\right)^2 + \frac12 f^2\sin^2\chi \, \partial_\mu \bm{n} \cdot \partial^\mu \bm{n} \nn
&= \frac12 \left[  \partial_\mu \bm{\varphi} \cdot \partial^\mu \bm{\varphi}
+  \frac{ \left( \bm{\varphi} \cdot \partial^\mu \bm{\varphi} \right)^2}{f^2 - \bm{\varphi \cdot \varphi}}\right]  \nn
&= \frac12 \partial_\mu \bm{\varphi} \cdot \partial^\mu \bm{\varphi}
+\frac1{2f^2}  \left( \bm{\varphi} \cdot \partial^\mu \bm{\varphi} \right)^2 \nn
&\qquad +\frac1{2f^4} \left( \bm{\varphi \cdot \varphi}\right) \left( \bm{\varphi} \cdot \partial^\mu \bm{\varphi} \right)^2  
+\ldots
\label{10}
\end{align}
in the two parametrizations. Since the scalar manifold has an $O(4)$ invariant fixed point $\bm{\phi}_0$, the $O(5)/O(4)$ model can be written as a SMEFT with $H$ given in terms of $\bm{\varphi}$ by Eq.~(\ref{1}),
\begin{align}
L 
&=  \partial_\mu H^\dagger \partial^\mu H
+\frac12 \frac{ \left[ \partial^\mu\left(H^\dagger H \right)\right]^2}{\left( f^2 - 2 H^\dagger H \right)}  \nn
&= \partial_\mu H^\dagger \partial^\mu H
-\frac1{2f^2} \left(H^\dagger H \right) \Box \left(H^\dagger H \right)  + \ldots
\label{10a}
\end{align}
up to terms of dimension six.

The Lagrangian Eq.~(\ref{10}) has four scalar degrees of freedom.
The angular part $\bm{n}$ has the three (eaten) Goldstone boson degrees of freedom, and the radial part $\chi$ has one degree of freedom. Both
$\bm{n}$ and $\bm{\varphi}$ transform as the fundamental of $O(4)$, and the group transformation law preserves the constraint $\bm{n\cdot n}=1$.
In the gauged case, one replaces $\partial_\mu \bm{n}$ and $\partial_\mu \bm{\varphi}$ by
\begin{align}
D_\mu \bm{n} &= \partial_\mu \bm{n} + \left( i g_L A_{L\mu} + i g_R A_{R\mu}\right) \bm{n}\,, \nn
D_\mu \bm{\varphi} &= \partial_\mu \bm{\varphi} + \left( i g_L A_{L\mu} + i g_R A_{R\mu}\right) \bm{\varphi}\,,
\end{align}
where the gauge generators are given in Eq.~(\ref{8p}). Note that $\chi$ is a gauge singlet,  so $D_\mu \chi = \partial_\mu \chi$.

Comparing with Fig.~\ref{fig:1}, the surface of the sphere $S^4$ in Fig.~\ref{fig:2} is the HEFT manifold $\m$, and the smaller red circle of radius $v$ in Fig.~\ref{fig:2} is the red curve marked $S^3$ in Fig.~\ref{fig:1}.  The four scalar fields
$\bm{\varphi}=f \sin \chi\, \bm{n}$ form the Higgs field $H$, as given in Eq.~(\ref{1}).
The SM gauge symmetry is obtained by gauging the $SU(2) \times U(1)$ subgroup of the unbroken $O(4)$. Since all points on the vacuum manifold $S^4$ are equivalent, we can choose the unbroken $O(4)$ group to be rotations about the $\phi_5$ axis, as shown in Fig.~\ref{fig:2}. 

If $O(5)$ symmetry is exact, then $\bm{\varphi}$ are exact Goldstone bosons, and there is no potential $V(\bm{\varphi})$. However, in composite Higgs models, one imagines that $\bm{\varphi}$ are approximate Goldstone bosons, and that some mechanism (not relevant for this paper) generates a potential that depends on the $O(4)$ invariant $\bm{\varphi \cdot \varphi}=f^2 \sin^2 \chi = 2 H^\dagger H$, i.e.\ the angle $\chi$ shown in Fig.~\ref{fig:2}. If this potential has a minimum not at the North pole, but at some small angle $\chi$, then the electroweak symmetry is broken by $v = f \sin \chi $, with $v \ll f$. The non-trivial task of composite Higgs models is to generate this small vacuum misalignment
angle.

The kinetic term defines the scalar metric, which is the induced metric on $S^4$,
\begin{align}
\rd s^2 &= f^2 \rd \chi^2 + f^2 \sin^2 \chi \left( \rd \bm{n} \cdot \rd \bm{n} \right)\,.
\label{11}
\end{align}
$\m=S^4$ is a four-dimensional maximally symmetric space of constant curvature, and the Riemann curvature tensor is
\begin{align}
R_{abcd}(\varphi) &= \frac{1}{f^2} \left( g_{ac}(\varphi) g_{bd}(\varphi) - g_{ad}(\varphi) g_{bc}(\varphi) \right)\,,
\label{12}
\end{align}
the Ricci tensor is
\begin{align}
R_{ab}(\varphi) &= \frac{1}{f^2} \left(N_\varphi-1\right) g_{ab}(\varphi)\,,
\label{13}
\end{align}
where $N_\varphi=4$ is the dimension of $\m$, and the scalar curvature is
\begin{align}
R(\varphi) &= \frac{1}{f^2} N_\varphi \left(N_\varphi-1\right) .
\label{14}
\end{align}

Comparing Eq.~(\ref{10}) with Eq.~(\ref{heft}), we see that
\begin{align}
\chi &= \chi_0 + \frac{h}{f}\,, & f^2 \sin^2 \chi &= v^2 F(h)^2\,,
\end{align}
and Eq.~(\ref{f0}) gives
\begin{align}
f^2 \sin^2 \chi_0 &= v^2\,,
\end{align}
so that
\begin{align}
F(h) &= \frac{f}{v} \sin \left[ \frac{h}{f} + \sin^{-1} \frac{v}{f} \right] \nn
&= \sqrt{\frac{f^2}{v^2}-1} \ \sin  \frac{h}{f}  + \cos \frac{h}{f}  \nn
&= 1 + \frac{h}{v} \sqrt{1-\frac{v^2}{f^2}} - \frac{h^2}{2f^2} +\ldots\,.
\end{align}
In the limit $f \to \infty$, the $O(5) \to O(4)$ model reduces to the SM,  the scalar manifold $\m$ becomes flat, and $F(h)$ reduces to the SM value Eq.~(\ref{fsm}).

The symmetry breaking pattern in the $O(5)$ model is
\begin{align}
O(5) \xrightarrow{\ \ f\ \ } O(4) \xrightarrow{\ \ v\ \ }  O(3)
\end{align}
which generates the inclusion
\begin{align}
O(5)/O(4)=S^4=\m \supset O(4)/O(3)= \msm\,.
\end{align}

%------------------------------------------------------------------------------
\section{The $O(4,1) \to O(4)$ model}\label{sec:h4}
%------------------------------------------------------------------------------

\begin{figure}
\centering
\includegraphics[]{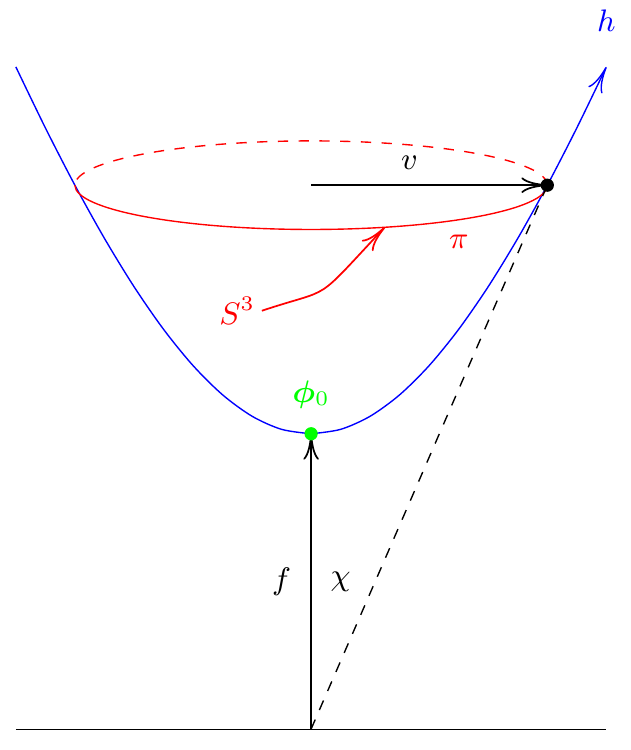}
\caption{\label{fig:4} The scalar manifold of the $O(4,1) \to O(4)$ sigma model is the hyperbolic space $\mathbb{H}^4$. The SM $SU(2) \times U(1)$ gauge group is a subgroup of the unbroken group $\h$ that leaves $\bm{\phi}_0$ invariant. The four-component field $\bm{\varphi}$ gets a VEV $v\ll f$, which breaks electroweak symmetry. $\mathbb{H}^4$ cannot be embedded in Euclidean space, so it is difficult to draw an accurate representation of the manifold. $\mathbb{H}^4$ has constant negative curvature, though this is not apparent from the figure.}
\end{figure}

We now consider a sigma model where $\m$ has negative curvature. Consider a five dimensional space with metric
\begin{align}
\rd s^2 &= \sum_{i=1}^4 \left(\rd \phi_i \right)^2 -  \left(\rd \phi_5 \right)^2
\label{26}
\end{align}
and the embedded surface $\m$ given by
\begin{align}
f^2 &=  \left( \phi_5 \right)^2- \sum_{i=1}^4 \left( \phi_i \right)^2 \,.
\end{align}
Choosing the branch $\phi_5>0$ gives the four dimensional hyperbolic space $\mathbb{H}^4$, which is a maximally symmetric space of negative curvature shown in Fig.~\ref{fig:4}. With coordinates
\begin{align}
\bm{\phi} &= \left[ \begin{array}{c} f \sinh \chi\, \bm{n} \\ f \cosh \chi \end{array}\right]
\label{31}
\end{align}
or
\begin{align}
\bm{\phi} &= \left[ \begin{array}{c} \bm{\varphi} \\[5pt] \sqrt{f^2+ \bm{\varphi \cdot \varphi}} \end{array}\right]\,,
\end{align}
%%%
%%% FIG 4
%%%
%%%
%%% END FIG 4
%%%

the metric on $\mathbb{H}^4$ is
\begin{align}
\rd s^2 &= \rd \bm{\phi} \cdot  \rd \bm{\phi} 
=f^2 \rd \chi^2 + f^2 \sinh^2 \chi \, \rd \bm{n} \cdot \rd \bm{n},
\end{align}
and the scalar kinetic energy term is
\begin{align}
L &=\frac12 f^2 \left(\partial_\mu \chi\right)^2 + \frac12 f^2\sinh^2\chi \, \partial_\mu \bm{n} \cdot \partial^\mu \bm{n} \nn
&= \frac12\left[ \partial_\mu \bm{\varphi} \cdot \partial^\mu \bm{\varphi}
- \frac{ \left( \bm{\varphi} \cdot \partial^\mu \bm{\varphi} \right)^2}{f^2 + \bm{\varphi \cdot \varphi}}\right]  \nn
&= \frac12 \partial_\mu \bm{\varphi} \cdot \partial^\mu \bm{\varphi}
-\frac1{2f^2}  \left( \bm{\varphi} \cdot \partial^\mu \bm{\varphi} \right)^2 \nn
&+\frac1{2f^4} \left( \bm{\varphi \cdot \varphi}\right) \left( \bm{\varphi} \cdot \partial^\mu \bm{\varphi} \right)^2  
+\ldots\,.
\label{15x}
\end{align}
$\mathbb{H}^4$ has an $O(4)$ invariant fixed point $\bm{\phi}_0$, so the $O(4,1)/O(4)$ model can be written as a SMEFT expansion, 
\begin{align}
L 
&=  \partial_\mu H^\dagger \partial^\mu H
-\frac12  \frac{ \left[ \partial^\mu\left(H^\dagger H \right)\right]^2}{\left( f^2 + 2 H^\dagger H \right)}  \nn
&= \partial_\mu H^\dagger \partial^\mu H
+\frac1{2f^2} \left(H^\dagger H \right) \Box \left(H^\dagger H \right)  + \ldots
\label{10b}
\end{align}
up to terms of dimension six.
 
$\mathbb{H}^4$ has curvature tensors
\begin{align}
R_{abcd}(\varphi) &= -\frac{1}{f^2} \left( g_{ac}(\varphi)  g_{bd}(\varphi)  - g_{ad}(\varphi)  g_{bc}(\varphi)  \right)\,, \nn
R_{ab}(\varphi)  &= -\frac{1}{f^2} \left( N_\varphi -1\right) g_{ab}(\varphi) \,, \nn
R(\varphi)  &= -\frac{1}{f^2} N_\varphi \left( N_\varphi-1\right)\, ,
\label{14n}
\end{align}
where $N_\varphi=4$.
The radial function is
\begin{align}
F(h) &= \frac{f}{v} \sinh \left[ \frac{h}{f} + \sinh^{-1} \frac{v}{f} \right] \nn
&= \sqrt{\frac{f^2}{v^2}+1} \ \sinh  \frac{h}{f}  + \cosh \frac{h}{f}  \nn
&= 1 + \frac{h}{v} \sqrt{1+\frac{v^2}{f^2}} + \frac{h^2}{2f^2} + \ldots\,.
\end{align}

\begin{figure}
\centering
\includegraphics[width=4cm]{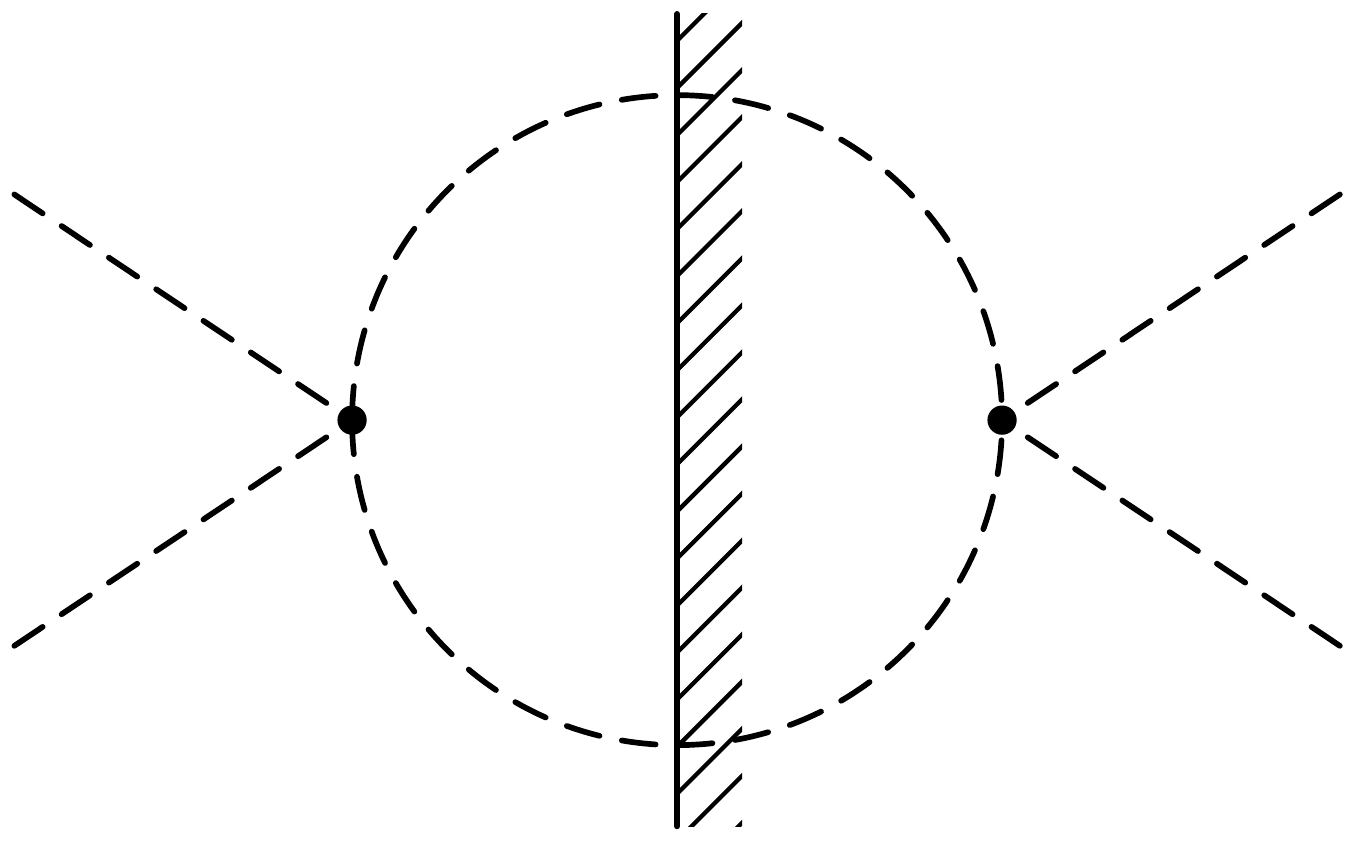}
\caption{\label{fig:3} The $\varphi \varphi$ forward scattering amplitude whose imaginary part gives the $\varphi \varphi \to \varphi \varphi$  total cross section. }
\end{figure}

Despite the minus sign in Eq.~(\ref{26}), the metric in Eq.~(\ref{15x}) is positive definite. The $\mathbb{H}^n$ sigma model defines a unitary theory of scalars, since the kinetic energy is positive definite. One way to see unitarity is to use the square root parameterization, and compare the $\mathbb{H}^n$ Lagrangian in Eq.~(\ref{15x}) with the $O(n)$ Lagrangian in Eq.~(\ref{10}).
The two differ by the replacement $f^2 \to -f^2$, so that the $4 \varphi$, $8\varphi$, etc.\ vertices have their signs flipped. The imaginary part of the $\varphi\varphi$ scattering graph in Fig.~\ref{fig:3} does not change sign, and remains equal to the $\varphi \varphi \to \varphi \varphi$ total cross section.
%%%
%%% BEGIN FIG 3
%%%
%%%
%%% END FIG 3
%%%

The Lagrangian Eq.~(\ref{15x}) has an $O(4,1)$ symmetry with 10 generators. Six of these are  generators of the $O(4)$ subgroup acting as rotations
on $\bm{\varphi}$. These generators are the same as $T^{ab}$ in Eq.~(\ref{5}) with $1 \le a  < b \le 4$. The remaining four generators are the boosts $B^a$, $1 \le a \le 4$,
\begin{align}
i \left[ B^{a}  \right]_{ij} &=  \delta_{ai} \delta_{5j} + \delta_{aj} \delta_{5i}\,
\end{align}
\begin{align}
i B^{1} &= \left[ \begin{array}{ccccc} 
0 & 0 & 0 & 0 & 1 \\ 
0 & 0 & 0 & 0 & 0 \\ 
0 & 0 & 0 & 0 & 0 \\ 
0 & 0 & 0 & 0 & 0 \\ 
1 & 0 & 0 & 0 & 0 \\ 
\end{array}\right],\ \text{etc.}
\end{align}
The symmetry group $O(4,1)$ is non-compact.
This structure should be familiar from the Lorentz group $SO(3,1)$. The space $\mathbb{H}^4$ is a homogeneous space, and we can choose the vacuum to be
\begin{align}
\vev{ \bm{\phi}_0} &= \left[ \begin{array}{c} 0 \\ 0 \\ 0 \\ 0 \\ f \end{array}\right]\,,
\label{24}
\end{align}
which breaks the non-compact $O(4,1)$ group down to the compact subgroup $O(4)$. Again, this should be familiar from the Lorentz group, where Eq.~(\ref{24}) is analogous to the momentum vector of a particle at rest, which breaks the boost generators but leaves the rotations unbroken.

We can now gauge some subgroup of the full symmetry group $O(4,1)$. At this stage, the non-compact nature of the symmetry group becomes important.
The gauge kinetic term is
\begin{align}
L &= -\frac14 G_{\mu \nu}^a G^{\mu \nu\, b} \mathcal{K}_{ab}
\end{align}
where $ \mathcal{K}_{ab}$ is the Killing form. For a compact Lie group, the generators are normalized so that $ \mathcal{K}_{ab}=\delta_{ab}$. However, for a non-compact group,
the Killing form  has some negative eigenvalues. In our example,
\begin{align}
\tr T^{ab} T^{cd} &= 2 \left( \delta_{ac} \delta_{bd} - \delta_{ad} \delta_{bc} \right)
\end{align}
for the $O(4)$ generators so that $\tr T^{ab} T^{ab} = 2 > 0$ (no sum on $a,b$), but
\begin{align}
\tr B^a B^b &= -2 \delta_{ab}
\end{align}
for the boost generators. Thus gauging the non-compact generators leads to a gauge boson kinetic energy with the wrong sign, so the theory is no longer unitary.
However, there is no problem if we only gauge a compact subgroup  of $O(4,1)$. For HEFT, we only need to gauge the $SU(2) \times U(1)$ compact subgroup of $O(4,1)$.  The gauged Lagrangian is given by replacing $\partial_\mu \bm{n}$ by $D_\mu \bm{n}$, as before.

The symmetry breaking pattern in the $O(4,1)$ model is
\begin{align}
O(4,1) \xrightarrow{\ \ f\ \ } O(4) \xrightarrow{\ \ v\ \ }  O(3)
\end{align}
which generates the inclusion
\begin{align}
O(4,1)/O(4)=\mathbb{H}^4=\m \supset O(4)/O(3)= \msm\,.
\end{align}

We can thus construct a HEFT where $\m$ has negative curvature, by gauging a $SU(2) \times U(1)$ subgroup of $O(4)$. The SM gauge symmetry is unbroken at the scale $f$. As in the compact $O(5)$ model of the previous section, one then needs to construct a vacuum misalignment mechanism where the HEFT field
$\bm{\varphi}=f \sinh \chi\, \bm{n}$ develops a VEV $v$, which breaks the SM gauge group. The unbroken symmetry group of the misaligned vacuum is a boosted version of $O(4)$, and is also compact.

%------------------------------------------------------------------------------
\section{Experimental Consequences}\label{sec:exp}
%------------------------------------------------------------------------------

In Ref.~\cite{Alonso:2015fsp}, we showed that gauge boson and Higgs boson scattering cross sections were related to the curvature of the HEFT manifold $\m$. The curvature functions defined in Ref.~\cite{Alonso:2015fsp} are
\begin{align}
\mathfrak{R}_4(h) &= \frac{f^2}{v^2} \sin^4 \left[ \frac{h}{f} + \sin^{-1} \frac{v}{f} \right] ,\nn
\mathfrak{R}_{2h}(h) &= \frac{1}{2} \sin^2 \left[ \frac{h}{f} + \sin^{-1} \frac{v}{f} \right] ,
\end{align}
for $O(5)$,
\begin{align}
\mathfrak{R}_4(h) &= -\frac{f^2}{v^2} \sinh^4 \left[ \frac{h}{f} + \sinh^{-1} \frac{v}{f} \right] ,\nn 
\mathfrak{R}_{2h}(h) &= -\frac{1}{2} \sinh^2 \left[ \frac{h}{f} + \sinh^{-1} \frac{v}{f} \right] .
\end{align}
for $O(4,1)$ and $\mathfrak{R}_i(h)=0$ for the SM. The curvature constants $\mathfrak{r}_i \equiv \mathfrak{R}_i(0)$ of Ref.~\cite{Alonso:2015fsp} are
\begin{align}
\mathfrak{r}_4 &= \kappa \frac{v^2}{f^2} \,, & \mathfrak{r}_{2h} &=  \kappa  \frac{v^2}{2f^2} \,,
\label{45}
\end{align}
where $\kappa=+1,-1, 0$ for positive, negative, and zero curvature, i.e.\ for the $O(5)$ model, $O(4,1)$ model, and SM, respectively.

In general, if $\g$ is compact, $\g/\h$ has non-negative sectional curvatures. The sectional curvature $K(X,Y)$ is defined
as
\begin{align}
K(X,Y) &\equiv \frac{ R_{abcd} X^a Y^b X^c Y^d}{ (g_{ac}g_{bd}-g_{ad}g_{bc}) X^a Y^b X^c Y^d} \nn
&= \frac{ \ip{R(X,Y)Y}{X}}{\ip{X}{X}\ip{Y}{Y}-\ip{X}{Y}\ip{X}{Y}}
\end{align}
for any two linearly independent tangent vectors $X$, $Y$, where $\ip{*}{*}$ is the inner product w.r.t.\ the metric $g_{ab}$.  Choosing $X,Y$ in the Goldstone boson directions and using the expression for $R_{abcd}$ in Ref.~\cite{Alonso:2015fsp} gives
\begin{align}
K(X_\varphi,Y_\varphi) &= \frac{1}{v^2} \mathfrak{R}_4(h),
\end{align}
so that $\mathfrak{R}_4(h) \ge 0$ which implies  $\mathfrak{r}_{4} \ge 0$. Choosing $X$ in the Goldstone boson direction and $Y$ in the Higgs direction gives
\begin{align}
K(X_\varphi,Y_h) &= \frac{2}{v^2}  \mathfrak{R}_{2h}(h),
\end{align}
so that $\mathfrak{R}_{2h}(h)\ge 0$ which implies $\mathfrak{r}_{2h} \ge 0$.

For  maximally symmetric spaces such as $S^n$ and $\mathbb{H}^n$, the sectional curvature is independent of $X,Y$.
Since the $O(5)/O(4)$ sigma model is based on a compact Lie group, the sectional curvatures are non-negative, and $\mathfrak{r}_{4} \ge 0$, $\mathfrak{r}_{2h} \ge 0$. For $O(5,1)/O(4)$,
$\mathfrak{r}_{4}$, $\mathfrak{r}_{2h}$  are \emph{negative}.

The HEFT $S$-parameter contribution
\begin{align}
\Delta S=\frac{1}{12\pi}\mathfrak{r}_{4}\log\left(\frac{\Lambda^2}{M_Z^2}\right)\,,
\end{align}
and the scattering amplitude of longitudinal $W$-bosons $W_L$ and Higgs bosons at high energy~\cite{Barbieri:2007bh,Contino:2010rs,Panico:2015jxa}
\begin{align}
\mathcal{A}\left( W_L W_L\to W_L W_L \right) &=-4\lambda + \frac{s+t}{v^2}\,\mathfrak{r}_4\, ,\nn
\mathcal{A}\left( W_L W_L\to h h \right) &= 2\lambda -\frac{2s}{v^2}\,\mathfrak{r}_{2h}\,.
\label{50}
\end{align}
depend on $\mathfrak{r}_4$ and $\mathfrak{r}_{2h}$. 
In particular, the sign of the new physics contribution depends on the curvature of the manifold.\footnote{In Eq.~(\ref{50}), the first term is the SM amplitude in the limit $s \gg M_H^2 \gg M_W^2$.}

Composite Higgs models considered in the literature are based on  
a compact group $\g$, and so the new physics contribution interferes destructively with the SM contribution. We have explicitly computed $\mathfrak{r}_4$ and $\mathfrak{r}_{2h}$ for the $O(5)$ theory in Eq.~(\ref{45}) and shown they are positive. While the values may change depending on the group structure of the $\g/\h$ sigma model, they remain non-negative for all theories based on a compact group $\g$.
For the negatively curved space considered here, $\mathfrak{r}_4$ and $\mathfrak{r}_{2h}$ are negative, and the interference is constructive.

Measuring deviations from the SM is a way of probing composite models, and gives direct information on the curvature of $\m$.
Experimental measurements based on perturbation theory calculations only probe the manifold in a neighbourhood of the vacuum (the black dot in Figs.~\ref{fig:1},\ref{fig:2},\ref{fig:4}). Since the $S$-matrix only depends on coordinate invariant properties of the manifold, the leading quantity which can be measured is the local curvature which is proportional to $1/f^2$, where $f$ is the new physics scale. Higher order corrections can depend on curvature gradients or higher powers of the curvature, and are suppressed by additional powers of $1/f$.

%------------------------------------------------------------------------------
\section{Vacuum Misalignment}\label{sec:align}
%------------------------------------------------------------------------------

The HEFT sigma models as described so far have exact Goldstone bosons, so that all points on $\m$ have the same energy. To have vacuum misalignment, it is necessary to generate a potential which breaks the exact $\g$ symmetry, so that the Goldstone bosons develop a small mass, and the minimum of  the potential is at a small vacuum misalignment angle $\chi$. Finding a suitable vacuum alignment mechanism is a difficult problem.

\begin{figure}
\includegraphics[width=3.5cm]{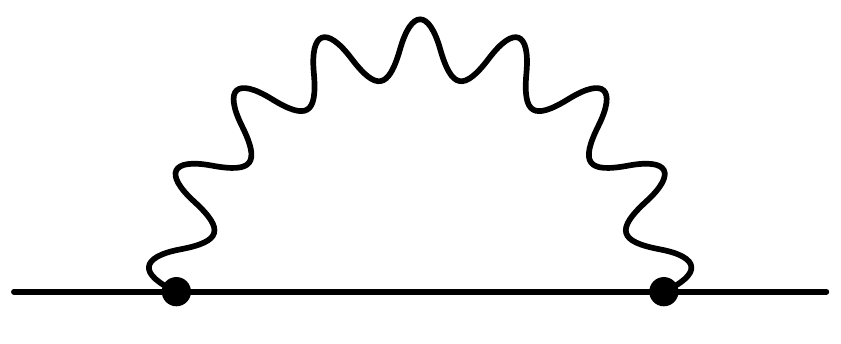}
\caption{\label{fig:qcd} Gauge contribution to pseudo-Goldstone boson masses.}
\end{figure}

One mechanism for mass generation is from gauge interactions, since only a subgroup of the original symmetry group $\g$ has been gauged. The same mechanism is responsible for the $\pi^+-\pi^0$ mass difference in QCD. Graphs such as Fig.~\ref{fig:qcd}
%%%
%%% - BEGIN FIG
%%%
%%%
%%% - END FIG
%%%
generate a mass difference, with a naive dimensional analysis~\cite{Manohar:1983md} estimate
\begin{align}
m^2_{\pi^+} - m^2_{\pi^0} \sim \frac{\alpha}{4\pi} \Lambda^2
\label{46}
\end{align}
which gives $\Delta m \sim 3$~MeV vs.\ the experimental value of 4.6~MeV in QCD.

In both the $O(5) \to O(4)$ and $O(4,1) \to O(4)$ sigma models in which the $SU(2) \times U(1)$ subgroup of $O(4)$ is gauged, one generates terms of the form
\begin{align}
\frac{g^2_2}{4\pi} \Lambda^2 \bm{\phi}^T T^a T^a \bm{\phi} &= \frac{3 \alpha}{4 \pi \sin^2 \theta_W} \Lambda^2 f^2 s^2(\chi)   \nn
\frac{g^2_1}{4\pi} \Lambda^2 \bm{\phi}^T YY \bm{\phi} &=  \frac{ \alpha}{4 \pi \cos^2 \theta_W} \Lambda^2  f^2 s^2(\chi)
\end{align}
at one loop order, where $s^2(\chi)=\sin^2\chi$, $\sinh^2 \chi$ or $\chi^2$ for positive, negative, and zero curvature, respectively. The gauge interactions give an effective potential
\begin{align}
V(\chi) &= c \,s^2(\chi)
\label{48}
\end{align}
where $c>0$, using the sign of the $\pi^+-\pi^0$ mass difference in Eq.~(\ref{46}). Eq.~(\ref{48}) has a minimum at $\chi=0$, which does not break electroweak symmetry. Various ideas have been proposed to solve the vacuum misalignment problem, including gauging an additional axial $U(1)_A$~\cite{Banks:1984gj}, or using top-quark loops~\cite{Agashe:2004rs,Contino:2010rs,Panico:2015jxa} to drive vacuum misalignment. Gauging an additional $U(1)_A$ requires extending the sigma model so that $U(1)_A$ is still part of a compact group.

Typically, the potential generated in $\g/\h$ models is of the form
\begin{align}
V(\chi) &= \alpha\, \cos \chi + \beta  \,\sin^2\chi
\label{49}
\end{align}
which breaks  electroweak symmetry if $\alpha - 2 \beta>0 $, $\beta < 0$, $\abs{\alpha/(2\beta)}<1$. For the negatively curved case, the effective potential is of the form
\begin{align}
V(\chi) &= \alpha\, \cosh \chi + \beta  \,\sinh^2\chi
\label{50V}
\end{align}
which breaks  electroweak symmetry if $2 \beta +\alpha < 0 $, $\beta>0$. The gauge contribution to $\beta$ should be positive in both cases, since the gauge bosons live in the compact part of the group. The top-quark scenario produces different values for $\alpha,\beta$ in the elliptic and hyperbolic cases, and a detailed analysis is needed to see if the electroweak symmetry is broken. Since $\chi$ is no longer periodic, it is also necessary
to check that the potential $V(\chi)$ is bounded from below. Some details of the $O(4,1)$ spinor algebra needed for the computation of the top-quark contribution to $V(\chi)$ are given in Appendix~\ref{app}.

%------------------------------------------------------------------------------
\section{Conclusions}\label{sec:concl}
%------------------------------------------------------------------------------

We have given a simple example of a sigma model with negative curvature based on a hyperbolic space. Deviations in the Higgs boson and longitudinal gauge boson scattering cross sections from their Standard Model values depend on the curvature, and have  opposite sign from the usual $\g/\h$ case of positive curvature. Thus, one can directly measure the curvature of the HEFT scalar space experimentally. $\g/\h$ sigma models  typically arise from breaking a compact flavor symmetry group $\g$ in a strongly interacting theory.  In this case $\g/\h$ has non-negative sectional curvatures, so that $\mathfrak{r}_4 \ge 0 $ and $\mathfrak{r}_{2h} \ge 0$. A detailed scenario that produces an example like the type discussed here requires the dynamics to produce a low-energy theory with a hyperbolic, rather than elliptic constraint. This scenario could occur in theories where the scalar manifold is complexified, as happens in supersymmetric theories. It would be interesting to study negatively curved sigma models in more detail to investigate unitarity and vacuum misalignment further.

\acknowledgments

We would like to thank R.~Contino and S.~Rychkov for helpful discussions.
This work was supported in part by grants from the Simons Foundation (\#340282 to Elizabeth Jenkins and \#340281 to Aneesh Manohar), and by DOE grant DE-SC0009919.

\begin{appendix}
%------------------------------------------------------------------------------
\section{$O(4,1)$ Spinors}\label{app}
%------------------------------------------------------------------------------

We briefly review some aspects of the $O(4,1)$ spinor representation needed for computing the top-quark induced potential. 
The results follow by analogy with known results on the relation between Euclidean and Minkowski space Dirac spinors for the Lorentz group.

The $SO(5)$ Clifford algebra is
\begin{align}
\left\{\Gamma^a,\Gamma^b\right\} &= 2 \eta^{ab}, & \eta^{ab} &= \mathop{\text{diag}}(1,1,1,1,-1) \,.
\end{align}
The $\Gamma$ matrices are
\begin{align}
\Gamma^{i} &= \left[ \begin{array}{cc} 0 & \sigma^i \\ \sigma^i & 0 
\end{array}\right], & i&=1,2,3 \nn
\Gamma^{4} &= \left[ \begin{array}{cc} 0 & -i \\ i & 0 
\end{array}\right] ,&
\Gamma^5 &= \left[ \begin{array}{cc} i & 0 \\ 0 & -i
\end{array}\right]\,.
\end{align}
The generators in the spinor representation are
\begin{align}
M^{[ij]} &= \frac{1}{4i} \left[\Gamma^i,\Gamma^j\right]\,.
\end{align}

If $\psi$ transforms as a spinor, 
\begin{align}
\psi &\to D(g)\, \psi\,,
\label{spin}
\end{align}
then
\begin{align}
\overline \psi \equiv \psi^\dagger \, \Gamma^5
\end{align}
transforms as the inverse,
\begin{align}
\overline \psi &\to \overline \psi \, D(g)^{-1}\,.
\end{align}

The analog of $\Gamma^i \Sigma^i$ in Refs.~\cite{Agashe:2004rs,Contino:2010rs} that enters the top-quark contribution to the
effective potential is
\begin{align}
\Gamma^5 \phi^i \Gamma_i 
&=f \left[ \begin{array}{cc}  \cosh \chi &  i \left( \bm{\sigma \cdot n} -i  n^4 \right) 
\sinh \chi \\   -i \left(\bm{\sigma \cdot n} + i n^4 \right) \sinh \chi  &  \cosh \chi
\end{array}\right] 
\end{align}
where $\phi^i$ is given in Eq.~(\ref{31}). If $\psi$, $\chi$ are spinors transforming as in Eq.~(\ref{spin}), then
\begin{align}
 \chi^\dagger \Gamma^5 \phi^i \Gamma_i \psi
\end{align}
is an invariant. The rest of the toq-quark calculation proceeds as in Refs.~\cite{Agashe:2004rs,Contino:2010rs}.

\end{appendix}

%----------------------------------------------------------------------

\bibliography{Higgs}

\end{document}